\documentclass[submission,copyright,fleqn]{eptcs}
\providecommand{\event}{SSV 2012}

\usepackage{amsmath,amssymb}
\usepackage{todonotes,comment}
\usepackage{graphicx}
\usepackage{subfig,wrapfig}
\usepackage{kivsyntax}
\usepackage{hyperref}

\title{A Formal Model of a Virtual Filesystem Switch}

\author{     Gidon Ernst
        \quad Gerhard Schellhorn
        \quad Dominik Haneberg
        \quad J\"org Pf\"ahler
        \quad Wolfgang Reif
\institute{Institute for Software and Systems Engineering, University of Augsburg, Germany}
        \email{\{ernst,schellhorn,haneberg,joerg.pfaehler,reif\}@informatik.uni-augsburg.de}}

\def\titlerunning{A Formal Model of a Virtual Filesystem Switch}
\def\authorrunning{G. Ernst, G. Schellhorn, D. Haneberg, J. Pf\"ahler \& W. Reif}

\renewcommand{\vec}[1]{\underline{#1}}
\renewcommand{\phi}{\varphi}

\newcommand{\Sec}[1]{Sec.~\ref{sec:#1}}
\newcommand{\Fig}[1]{Fig.~\ref{fig:#1}}
\newcommand{\Eqn}[1]{(\ref{eqn:#1})}

\newcommand{\Section}[1]{Section~\ref{sec:#1}}
\newcommand{\Figure}[1]{Figure~\ref{fig:#1}}

\mathchardef\mhyphen="2D

\def\p#1{\mathrel{\ooalign{\hfil$\mapstochar\mkern 5mu$\hfil\cr$#1$}}}
\newcommand{\pto}{\p\to}

\newcommand{\oh}      {\mathit{oh}}

\newcommand{\dirs}    {\mathit{dirs}}
\newcommand{\files}   {\mathit{files}}

\newcommand{\fd}      {\mathit{fd}}

\newcommand{\str}     {\mathit{str}}
\newcommand{\ino}     {\mathit{ino}}

\newcommand{\INV}     {\mathit{INV}}
\newcommand{\vpages}  {\mathit{pages}}

\newcommand{\N}       {\mathbb{N}}
\newcommand{\Bool}    {\mathbb{B}}

\newcommand{\Seq}     {\mathit{Seq}}
\newcommand{\Array}   {\mathit{Array}}
\newcommand{\Buffer}  {\mathit{Buffer}}

\newcommand{\Byte}    {\mathit{Byte}}
\newcommand{\Page}    {\mathit{Page}}

\newcommand{\String}  {\mathit{String}}

\newcommand{\Path}    {\mathit{Path}}
\newcommand{\Mode}    {\mathit{Mode}}
\newcommand{\Error}   {\mathit{Error}}
\newcommand{\Meta}    {\mathit{Meta}}
\newcommand{\User}    {\mathit{User}}
\newcommand{\File}    {\mathit{File}}
\newcommand{\Dir}     {\mathit{Dir}}
\newcommand{\Ino}     {\mathit{Ino}}
\newcommand{\Inode}   {\mathit{Inode}}
\newcommand{\Dentry}  {\mathit{Dentry}}
\newcommand{\Handle}  {\mathit{Handle}}

\newcommand{\mkfile}  {\mathtt{file}}
\newcommand{\mkinode} {\mathtt{inode}}
\newcommand{\mkdir}   {\mathtt{dir}}
\newcommand{\mkhandle}{\mathtt{handle}}
\newcommand{\mkdentry}{\mathtt{dentry}}
\newcommand{\mknegdentry}{\mathtt{negdentry}}

\newcommand{\ROOTINO} {\mathtt{ROOT\_INO}}
\newcommand{\PAGESIZE}{\mathtt{PAGE\_SIZE}}
\newcommand{\EIO}     {\mathtt{EIO}}

\newcommand{\ESUCCESS}{\mathtt{ESUCCESS}}

\newcommand{\dom}     {\mathtt{dom}}

\newcommand{\links}   {\mathtt{links}}

\newcommand{\isopen}  {\mathtt{is\mhyphen open}}

\newcommand{\px}      {\mathtt{px}}
\newcommand{\pr}      {\mathtt{pr}}
\newcommand{\pw}      {\mathtt{pw}}
\newcommand{\content} {\mathtt{content}}

\newcommand{\meta}    {\mathtt{meta}}
\newcommand{\size}    {\mathtt{size}}
\newcommand{\pages}   {\mathtt{pages}}
\newcommand{\entries} {\mathtt{entries}}
\newcommand{\name}    {\mathtt{name}}
\newcommand{\sino}    {\mathtt{ino}}
\newcommand{\spos}    {\mathtt{pos}}
\newcommand{\smode}   {\mathtt{mode}}
\newcommand{\isdir}   {\mathtt{isdir}}
\newcommand{\nlink}   {\mathtt{nlink}}
\newcommand{\target}  {\mathtt{target}}

\begin{document}
\maketitle

\begin{abstract}
This work presents a formal model that is part of our effort to construct a
verified file system for Flash memory.
To modularize the verification we factor out generic aspects into a common
component that is inspired by the Linux Virtual Filesystem Switch (VFS)
and provides POSIX compatible operations.
It relies on an abstract specification of its internal interface
to concrete file system implementations (AFS).
We proved that preconditions of AFS are respected and that the state is kept consistent.
The model can be made executable and mounted into the Linux directory tree using FUSE.
\end{abstract}

\section{Introduction}
\label{sec:introduction}

The popularity of Flash memory as a storage technology has been increasing
constantly over the last years. Flash memory offers a couple of advantages
compared to magnetic storage: It is less susceptible to mechanical shock,
consumes less energy and offers higher speed when reading data.
However, Flash memory can only be written sequentially, and memory cells
must be erased in rather large blocks before they can be written again.

Two approaches exist to deal with these special characteristics:
Standard file systems can be used if the hardware has a built-in 
\emph{Flash translation layer} (FTL) that emulates behavior of magnetic storage.
USB drives and Solid State Disks fall into this category.
In contrast, Flash file systems (FFS for short) are specifically designed to work
on ``raw-flash''. FFS are commonly used in embedded systems such as home
routers, example implementations are YAFFS and JFFS.
More recently, \emph{UBIFS} \cite{ubifs-whitepaper08} has become part of the
Linux kernel and represents the state of the art.
An FFS can in principle be more efficient than the combination of FTL and a
traditional file system.

Flash memory is beginning to be used in safety-critical applications,
leading to high costs of failures and correspondingly to a demand for high
reliability of the file system implementation.
As an example, an error in the software access to the Flash store of the Mars
Exploration Rover ``Spirit'' already had nearly disastrous consequences
\cite{MarsFlashAnomaly05}.
As a response, Joshi and Holzmann \cite{Joshi-Holzmann07} from the NASA JPL
proposed in 2007 the verification of a Flash file system as a pilot
project of Hoare's Verification Grand Challenge \cite{Hoa03}
and for use in future missions.

We are developing such a verified Flash file system as an implementation of the
POSIX file system interface \cite{web:POSIX}, using UBIFS as a blueprint.
Our goal is that it can either be used \emph{stand-alone} or in \emph{Linux}.

The effort is structured into layers that are connected by refinement,
corresponding to the various logical parts of the file system.
The top level is an abstract formal model of the file system interface as
defined in the POSIX standard.
It serves as the specification of the functional requirements, i.e., what it
means to create/remove a file/directory and how the contents of files are
accessed.

The POSIX interface addresses files and directories by \emph{paths} and views
files as a linear sequence of bytes. Such high-level concepts are typically
mapped to a more efficient data representation in the file system,
in particular a graph structure.
In Linux, this mapping is realized by the \emph{Virtual Filesystem Switch} (VFS).
It implements many generic operations that are common to all file systems, e.g.,
permission checks and management of open file handles.
VFS relies on concrete file systems -- such as UBIFS -- to provide lower-level
operations. To that purpose, VFS defines an internal interface.
The advantage of this scheme is \emph{separation of concerns} and code reuse.

To achieve a fully verified POSIX compatible file system we have to provide our
own implementation of the VFS functionality.
Therefore, we define a formal specification of the main operations of the
internal interface of the Linux VFS, called AFS%
\footnote{Not to be confused with the Andrew File System}%
(for abstract file system) in the following.

\begin{figure}
    \parbox{0.39\textwidth}{
        \centering
        \includegraphics[height=5cm]{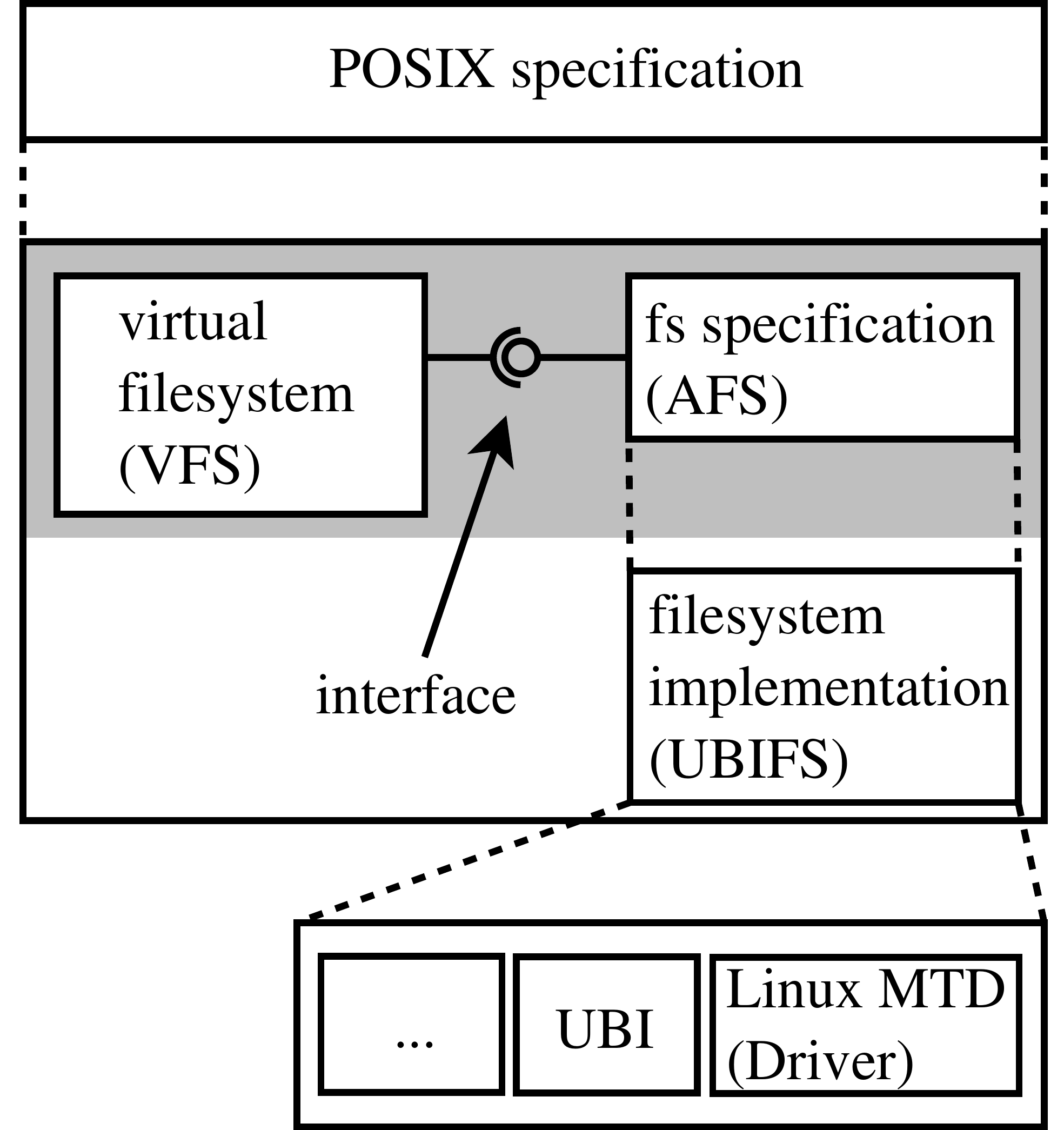}
        \caption{Components and Structure}
        \label{fig:structure}
    }
    \parbox{0.3\textwidth}{
        \centering
        \vspace{1cm} 
        \includegraphics[height=4cm]{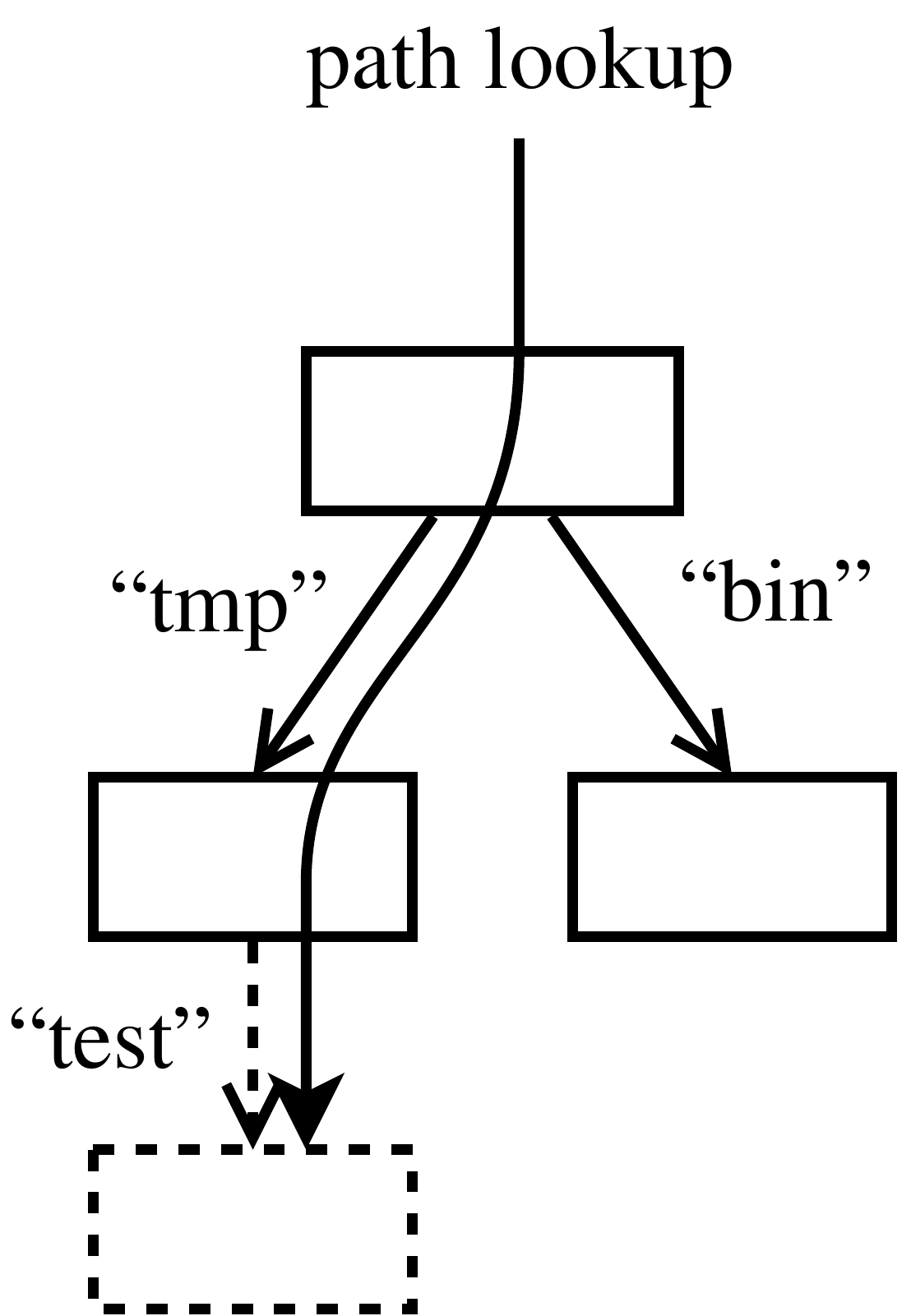}
        \caption{File system graph}
        \label{fig:graph}
    }
    \parbox{0.3\textwidth}{
        \centering
        \vspace{1cm} 
        \includegraphics[height=4cm]{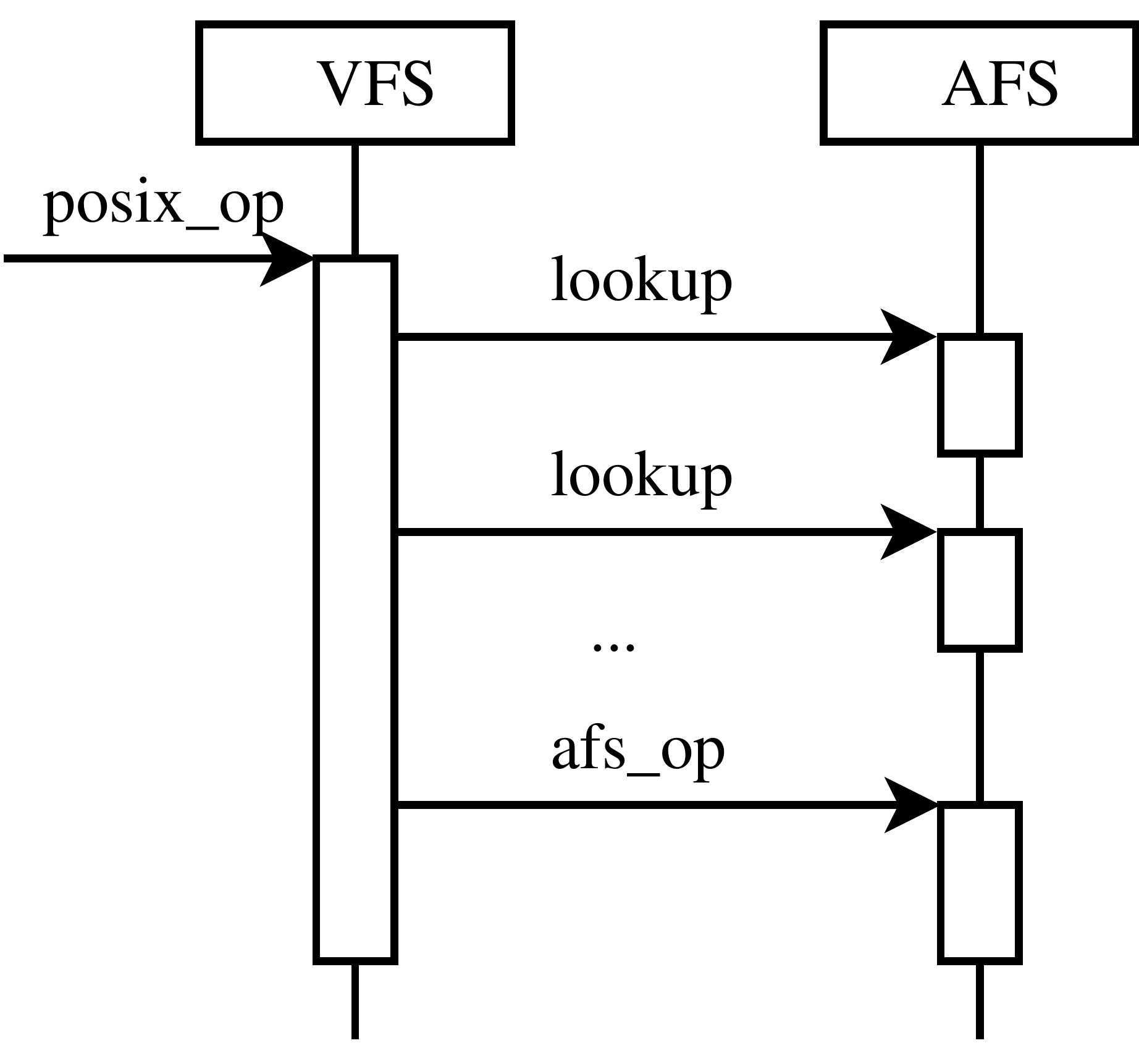}
        \caption{VFS/AFS interplay}
        \label{fig:sequence}
    }
\end{figure}

\Figure{structure} visualizes the different components displayed as boxes.
Functional correctness is established by several nested \emph{refinements}
(or simulation relations) that are graphically shown as dotted lines.
In particular, the a proof of the topmost refinement from POSIX to VFS+AFS
implies the correctness of the model presented in this paper.

Note that AFS is refined to a Flash File System independently of VFS.
As a consequence, VFS on top of the concrete FFS instead of AFS also refines
POSIX, i.e., functional correctness propagates in a compositional fashion.

Previously published is our formalization of the core concepts of UBIFS
\cite{FlashFM09} which deals with keeping the index data structures consistent.
A POSIX model and the two refinements from POSIX to VFS and from AFS to UBIFS
are ongoing work.
We also work on the formalization of other layers, including UBI that takes the
role of an FTL.

Our specification language is \emph{Abstract State Machines}
\cite{asmbook03}, which define operations over abstract data types.
We use algebraic specifications to axiomatize data types, and our interactive
theorem prover KIV \cite{RSSB98dfgife} to verify properties.

\noindent
In summary, the contribution of this work is:
\begin{itemize}
\item
A model of operations common to file systems, i.e., a Virtual Filesystem Switch,
that provides all essential POSIX file system interface.
Standard POSIX operations are broken down to a number of AFS operations.
Notably, linear access to file content is mapped to a sparse array of pages.
Besides that, we implement path lookup, access permission checks and management
of handles for open files. The model is thus very close to an implementation.

\item
The AFS model for file systems that can be plugged into the above VFS.
It abstracts details of concrete file system operations into a generic
specification and encodes the assumptions made by the VFS.
Thus, if a file system implementation respects the given contracts,
it can be used directly within our VFS.
In contrast to the VFS model, AFS remains as abstract as possible.

\item
We verified that AFS preconditions are respected by the VFS and we proved
several consistency invariants about the state. 
\end{itemize}

Furthermore, we have derived an executable Scala simulation from the models that
can be mounted directly into the Linux directory tree using the file system in
user space library (FUSE \cite{fuse}).
The simulation is used for testing and validates the models with respect to POSIX.
For reasons of space we do not present the full models in this paper
(AFS: $\sim$100 LOC, VFS: $\sim$500 LOC).
Together with the proofs of invariants and the code of the simulation they are
available online \cite{KIV-VFS-AFS-web12}.

The remainder of this paper is structured as follows:
\Section{vfs} describes the VFS data model and operations, focusing on
structural modifications of the file system.
\Section{afs} describes the AFS data types, operations and invariants.
We then turn to access of file content through file handles in \Sec{rw}.
\Section{related} discusses related work and \Sec{conclusion} concludes.

\section{The VFS Layer}
\label{sec:vfs}

This section shows, how the VFS layer realizes the top-level POSIX operations.
Conceptually, the file system consists of directories and files that
are organized hierarchically in a \emph{directed acyclic graph} with the toplevel
directory as root node, as visualized in \Fig{graph}.
File system objects are addressed by paths, which are sequences of names
concatenated by the separator \texttt{\char`/} (resp.  \texttt{\char`\\} on Windows).
Operations can be classified into structural modifications, such as
creating/deleting files, and content modification, such as reading/writing.

Creation of a new file will be used as running example throughout the paper.
The following C source code creates a file named ``test'' in
the top-level directory named ``tmp'', using the \texttt{creat(3)} operation:%
\begin{verbatim}
    int err = creat("/tmp/test", 0644);
\end{verbatim}
The given path is parsed, each segment but the last is looked up in the
directory tree (starting from the root) and finally a file with some access
permissions (here: 0644) is created.
The return value indicates success or a specific error condition.
The new subgraph arising from the \texttt{creat} operation is indicated by
dotted lines in \Fig{graph}.

The task of the VFS layer is to break down such high-level POSIX operations to
several calls of AFS operations. \Fig{sequence} visualizes a typical sequence
for structural operations like \texttt{creat}. In this case, it relies on three
operations provided by the file system implementation, namely

1) lookup of the target of a single edge in the graph

2) retrieve the access permissions at each node that is passed during path lookup

3) actual creation of the file

\noindent In Linux, the AFS interface is realized by a set of function pointers.
The file system specific create operations have the following signature:
\begin{verbatim}
    int (*create)(struct inode *dir, struct dentry *dent,
                  int mode, struct nameidata *nd);
\end{verbatim}

The first parameter, \texttt{dir}, points to some object representing the parent
directory of the new file (``/tmp'' in the example above).
The second parameter, \texttt{dent}, specifies the name of the new file.

The types of the formal parameters of the internal interface constitute the
\emph{VFS data model}. All file system queries and modifications are expressed
in terms of these data structures.

\subsection{Data Model}
\label{sec:vfs:data}

This section defines an algebraic specification of the VFS data model.
Three main data structures represent the file system graph:
\emph{inodes}, \emph{dentries} and \emph{pages}.
They are \emph{communication} data structures and do not necessarily reflect
the file system's runtime state and the on-disk data structures.

\emph{Inodes} (``Index Nodes'') correspond to the nodes of the graph, i.e.,
the files and directories. Inodes are uniquely identified by an inode
number $\ino : \Ino \simeq \N$ and store some associated information.
The sort $\Inode$ is formally defined as an algebraic data type:
\begin{align*}
\textbf{data } \Inode\
    =\ \mkinode(\sino : \Ino,\ \meta : \Meta,\ \isdir : \Bool,\
                \nlink : \N,\ \size : \N)
\end{align*}
It has one constructor $\mkinode$ that records the inode number ($\sino$),
some metadata ($\meta$),
whether it corresponds to a file or a directory ($\isdir$),
the number of hard-links (inbound edges, $\nlink$)
and the file size resp.\ the number of directory entries in case of a directory
inode ($\size$).
The abstract sort $\Meta$ is a placeholder for any further information
that is associated with inodes.
We postulate some selectors, to retrieve for example read, write and execute
permissions following the ideas given in \cite{Hesselink-Lali-FACS12}.
Constructor arguments are accessed by postfix selectors with the given names,
for example $\mathit{inode}.\size$ retrieves the size of an inode $\mathit{inode}$.

\emph{Dentries} (``Directory Entries'') correspond to the edges of the graph.
They relate a parent directory to its children and are labelled with the
respective \emph{file names}.
Dentries store a $\name$ and come in two flavors:
Normal dentries point to an existing file identified by the selector $\target$.
Negative dentries indicate that a file name is \emph{not} contained within a
directory -- they are used for example as return value of the lookup operation.
%
\begin{align*}
\textbf{data } \Dentry\
     =&\ \mkdentry(\name : \String,\ \target : \Ino)~~ |~~ \mknegdentry(\name : \String)
\end{align*}

The content of files is partitioned into uniformly sized \emph{pages}.
This has several advantages: The size of pages typically corresponds to the size
of a virtual memory page, enabling caching and memory-mapped input/output.
Furthermore, sparse files, i.e. files with large empty parts, can be represented
efficiently by the convention that non-present pages contain zeros only.
Formally, pages are arrays of bytes of a fixed length, specified by the
constant $\PAGESIZE$:
\begin{align*}
\textbf{type } \Page\ 
    =\ \Array[\PAGESIZE]\langle\Byte\rangle
\end{align*}

\subsection{Operations}
\label{sec:vfs:operations}

To continue the example, the signature of the create operation in our model is:
\begin{verbatim}
    vfs_create(path : Path, md : Meta; err : Error)
\end{verbatim}
where paths are sequences of strings (the separator is implicit) and errors are
given by an enumeration of possible error constants (where $\ESUCCESS$ denotes
``no error''):
\begin{align*}
\textbf{type } \Path  = \Seq\langle\String\rangle 
\qquad
\textbf{data } \Error = \ESUCCESS \mid \EIO \mid \ldots
\end{align*}

As a convention, we prefix VFS and AFS operations with \texttt{vfs\_} resp.\
\texttt{afs\_} to indicate that they are part of the formal model.
The semicolon in the parameter list separates input parameters from reference
parameters. Thus, assignments to \texttt{err} in the body of the operations are
visible to the caller.


VFS (and AFS) operations are defined by \emph{rules} of an
Abstract State Machine (ASM) \cite{asmbook03}.
The language features typical programming constructs such as parallel (function)
assignments, conditionals, loops, nondeterministic choice and recursive
procedures. ASM rules are executable, provided that the nondeterminism is
resolved somehow and the algebraic operations on data types are executable.

All VFS operations perform extensive error checks.
Similar to Hesselink and Lali \cite{Hesselink-Lali-FACS12}, we ensure that the
file system is guarded against unintended or malicious calls to operations.
Specifically, all operations are total (defined for all possible values of
input parameters) and either succeed or return an error \emph{without} modifying
the internal state.
Some implications of these checks manifest as preconditions in AFS
(and further refinements), as discussed in \Sec{afs:operations}.

\Figure{vfs-create} shows the ASM rules that realize the create operation in the
syntax of our specification language. The entry point is \texttt{vfs\_create}.
It receives the path to the new file and ensures that it is non-empty.
Subsequently, the helper routine \texttt{vfs\_walk} determines the parent
directory of the new file, identified by the inode number returned in
\texttt{ino}. The path walk itself performs multiple lookups of directory
entries by calling the AFS routine \texttt{afs\_lookup(ino; dent, err)}.
Informally, \texttt{afs\_lookup} checks, whether the name given by \texttt{dent}
is contained in the directory identified by \texttt{ino} and sets
\texttt{dent.target} to the inode number of the child or returns an error.
In the latter case, the whole operation is aborted.
The subroutine \texttt{vfs\_maycreate} ensures that the user has sufficient
permissions to create a file in the parent directory and that the name is not
already present.
Creation of the file is then delegated to \texttt{afs\_create(ino, md; dent, err)}
(shown in \Fig{afs-create}),
which allocates a new inode number and modifies its internal state.
The new inode number is returned in \texttt{dent.target}.

\begin{figure}[t]
    \centering
\begin{lstlisting}[style=kiv,language=kiv,mathescape=true,multicols=2]
vfs_create(path, md; err)
  if path = [] then
    err := EACCESS
  else let
    ino   = ROOTINO,
    dent  = negdentry(path.last)
    path  = parent(path)
  in vfs_walk(path; ino, err);
     if err = ESUCCESS then
        vfs_maycreate(ino; dent, err);
     if err = ESUCCESS then
       afs_create(ino, md; dent, err)

vfs_walk(path; ino, err)
  err := ESUCCESS;
  let path = path in
  while path != [] && err = ESUCCESS do
    vfs_maylookup(ino; err);
    if err = ESUCCESS then
    let dent = negdentry(path.first)
    in  afs_lookup(ino; dent, err);
        if err = ESUCCESS then
          ino := dent.target
          path := path.rest

$~$
\end{lstlisting}
    \caption{ASM rules of the VFS create operation}
    \label{fig:vfs-create}
\end{figure}

The formal model supports the following operations besides \texttt{create}:

The operation \texttt{mkdir} creates a new empty directory, conversely,
\texttt{rmdir} removes an existing directory, which must be empty.
The operation \texttt{link} introduces hard-links to existing files,
i.e., they introduce additional edges in the graph between existing nodes,
so that a file becomes accessible through multiple paths.
The converse operation is \texttt{unlink}, which also deletes the file on disk
when the last link is removed (and the last file handle is closed, see \Sec{evict}).
The most complex structural operation is \texttt{rename}
(AFS rule shown in \Fig{afs-rename}), which allows to change
the name and optionally the parent directory of a file or directory.
It performs two path walks and checks a couple of error conditions.


The operation \texttt{open} returns a \emph{file descriptor} through which the
content of the file can be accessed with \texttt{read}, \texttt{write}, and \texttt{seek}.
The operation \texttt{close} invalidates a descriptor and frees associated resources.
The size of a file can be changed with \texttt{truncate}, meta data is accessed
with \texttt{getattr} and \texttt{setattr}.
Finally, the operation \texttt{readdir} returns the filenames contained in a directory.

\section{The AFS Layer}
\label{sec:afs}

AFS is also realized as an Abstract State Machine, i.e., it is an
\emph{operational} specification of expected behavior of concrete file systems.
The design goal is to remain as abstract as possible.

\subsection{State}
\label{sec:afs:data}

AFS maintains an internal state, consisting of files and directories.
These are kept in two separate stores (partial functions), mapping inode numbers
to the respective objects; we use the symbol $\pto$ to denote partial function types.
\begin{align*}
\textbf{state vars }
    \quad
    \dirs : \Ino \pto \Dir,
    \quad
    \files : \Ino \pto \File
\qquad \text{where } \Ino \simeq \N
\end{align*}
The separation is motivated by the distinction into structural and content modifications:
the former will affect mainly $\dirs$ while the latter will affect only $\files$.
We expect this decision to facilitate the refinement proofs between the POSIX
layer and VFS. However, it comes at the cost of an extra disjointness
invariant (see \Sec{afs:invariants}).

Directory entries are \emph{contained} in the parent directory, likewise, pages
are contained in the file object they belong to:
\begin{align*}
\textbf{data } \Dir\: &= \mkdir(\meta: \Meta,\ \entries: \String \pto \Ino)         \\
\textbf{data } \File  &= \mkfile(\meta: \Meta,\ \size: \N,\ \pages: \N \pto \Page)
\end{align*}
Inode numbers $\ino \in (\dom(\dirs)\ \cup\ \dom(\files))$ are called
\emph{allocated}, they refer to valid directories resp.\ files.
We often omit the $\dom$-operator for brevity, as in $\ino \in \dirs$.
We write application of partial functions using square brackets, e.g.,
$\dirs[\ino]$ retrieves the directory identified by $\ino$.
Function update is written $\dirs[\ino] := d$ for a directory $d$.
We require explicit allocation of inode numbers for $\dirs$ and $\files$,
$\dirs \textrm{ +\,+ } \ino$ denotes the store $\dirs$ with an additional
mapping for $\ino$ (to an arbitrary directory).
Conversely, deallocation is written as $\dirs \textrm{ --\,-- } \ino$.
Empty stores are written as $\emptyset$.
Similar conventions apply to the stores $\files$, $\entries$ and $\pages$,
except that the latter two do not require explicit allocation.

\subsection{Operations}
\label{sec:afs:operations}

ASM rules in the AFS model have the form
\texttt{if {\slshape pre} then {\slshape actions}}
with a precondition \texttt{\slshape pre}.
These preconditions roughly correspond to the error checks performed by VFS.
We proved that all calls from VFS indeed establish the AFS preconditions.
In contrast to the axiomatic approach of Hoare-style contracts,
postconditions are implicitly given by the effect of the actions,
namely, the outputs and state transitions.

All structural operations, such as \texttt{vfs\_create}, have corresponding AFS counterparts.
\Figure{afs-create} shows the ASM rules of \texttt{afs\_lookup} and \texttt{afs\_create}.

The lookup operation tests whether a parent directory identified by the inode
number \texttt{pino} contains an entry with a specific name,
given by the (negative) directory entry \texttt{dent}.
It returns a positive directory entry that points to the child's inode number or
a negative directory entry and the error ``no entry''.
The test \texttt{pino $\in$ dirs} represents the precondition of the operation:
it may only be invoked with a valid parent inode number.

The create operation allocates a new inode number \texttt{ino}. It is added to
the $\entries$ slot of the parent directory under \texttt{dent.name}.
The new file object has the given metadata, a size of zero and no pages.
The operation requires that \texttt{pino} refers to a valid directory and
that the name is not already present.

\begin{figure}
    \centering
\begin{lstlisting}[style=kiv,language=kiv,mathescape=true,multicols=2]
afs_lookup(pino; dent, err)
if pino _in $~$dirs then
  if dent.name _in dirs[pino].entries
  then let
    ino = dirs[pino].entries[dent.name]
    in dent := dentry(dent.name, ino)
       err := ESUCCESS
  else dent := negdentry(dent.name)
       err := ENOENT



afs_create(pino, md; dent, err)
if   pino _in $~$dirs
   && dent.negdentry?
   && dent.name _notin dirs[pino].entries
then
  choose   ino with _not ino _in dirs
         && _not ino _in files && ino != 0 in
  dirs[pino].entries[dent.name] := ino
  files := files ++ ino
  files[ino] := file(0, md, _emp)
  dent := dentry(dent.name, ino)
  err := ESUCCESS
\end{lstlisting}
    \caption{ASM rules of the AFS create and lookup operations}
    \label{fig:afs-create}
\end{figure}

The AFS rename operation is shown in \Fig{afs-rename}.
It takes two inode numbers and two directory entries.
The file or directory is simply renamed if $\texttt{oldino} = \texttt{newino}$.
Otherwise, it is additionally moved into a different parent directory.
It is possible to overwrite an existing destination, if it has the same type
(file resp. directory) and in the latter case the destination is empty.
Furthermore there are some consistency conditions on both dentries.
The state transition consists of two directory modifications.
Note that sequential composition (emphasized by curly braces)
in the given order is significant,
since the two statements might affect the same parent directory.

\begin{figure}
    \centering
\begin{lstlisting}[style=kiv,language=kiv,mathescape=true]
afs_rename(oldino, newino; olddent, newdent)
  if   oldino _in dirs
     && newino _in dirs
     && olddent.target _in dirs[oldino].entries
     && (  olddent.ino _in files && newdent.ino _in files
        || olddent.ino _in dirs  && newdent.ino _in dirs )
     && (newdent.ino _in dirs -> dirs[newdent.ino].entries = _emp)
     && ;; whether olddent/newdent must be positive/negative,
       ;;         and their .target inodes are allocated
  then
    { dirs[oldino].entries := dirs[oldino].entries -- olddent.target;
      dirs[newino].entries[newdent.target] := olddent.target }
    olddent := negdentry(olddent.target)
    newdent := dentry(newdent.target, olddent.target)
    err := ESUCCESS
\end{lstlisting}
    \caption{ASM rules of the AFS rename operations}
    \label{fig:afs-rename}
\end{figure}

The remaining non-structural AFS operations are:
The pair of operations \texttt{afs\_readinode} and \texttt{afs\_writeinode}
construct $\Inode$ instances and write back changes.
Similarly, \texttt{afs\_readpage} and \texttt{afs\_writepage} read and write
whole pages. The operation \texttt{afs\_evict} deletes unreferenced files.
The operations \texttt{afs\_readdir} and \texttt{afs\_truncate} complete the list.

\subsection{Nondeterministic Errors}
\label{sec:afs:errors}

AFS operations assume infinite, perfect storage: allocation always succeeds and
$\dirs$ and $\files$ can be accessed reliably. However, transient and persistent
failures are quite common with real Flash hardware; allocation may fail due to
insufficient memory, and the Flash device can be full.
Since the AFS model is too abstract to capture when exactly such errors arise,
most AFS operations nondeterministically fail with an error code selected from a
set of low-level errors
(the exception is 
\texttt{afs\_evict}, see \Sec{evict}).
This is achieved by modifying the body of operations as follows:
\begin{lstlisting}[style=kivbig,language=kiv,mathescape=true]
{ choose e with e _in {EIO,...} in err := e } or { $\text{\slshape\ttfamily normal code}$ }
\end{lstlisting}

\subsection{Invariants}
\label{sec:afs:invariants}

We have proved the following invariants on the AFS state:
inode numbers of files and directories are disjoint, never $0$ and there is a
root inode \Eqn{afs:ino}.
Invariant \Eqn{afs:closed} states closure under lookup: following a directory
entry leads to an allocated inode number.
Directories have at most one parent \Eqn{afs:links} and
there are no superfluous pages beyond the file size \Eqn{afs:pages}.
For all $\ino$, $\str$, $n$:
\begin{gather}
0 \not\in \dirs     \quad \text{ and } \quad
0 \not\in \files    \quad \text{ and } \quad
\ROOTINO \in \dirs  \quad \text{ and } \quad
\dirs \cap \files = \emptyset
\label{eqn:afs:ino}
\\
\ino \in \dirs \land \str \in \dirs[\ino].\entries ~ \to ~ \dirs[\ino].\entries[\str] \in (\dirs \cup \files)
\label{eqn:afs:closed}
\\
\ino \in \dirs ~ \to ~ \#~\links(\ino, \dirs) \le 1
\label{eqn:afs:links}
\\
\ino \in \files \land n \in \files[\ino].\pages ~ \to ~ n * \PAGESIZE < \files[\ino].\size
\label{eqn:afs:pages}
\\
\text{The last page of a file (if present) contains zeros in the part that is outside the file size}
\label{eqn:afs:zero}
\end{gather}
The proofs are not difficult, a couple of helper lemmas lead to high automation.
The verification crucially relies on the AFS preconditions. For example:
\begin{itemize}
\item
    The \texttt{link} operation requires the target to be a file,
    otherwise, invariant \Eqn{afs:links} may be violated.
\item
    The \texttt{rmdir} operation requires that the path is not empty,
    so that the root directory is never deleted (invariant \Eqn{afs:ino}).
\end{itemize}

\section{File Access and File Handles}
\label{sec:rw}

The external POSIX view of file content is a sequence of bytes that is accessed
indirectly through \emph{file descriptors}, passed to the operations read and write.
An example write to the file ``/tmp/test'' using the C interface is:
\begin{verbatim}
     int fd = open("/tmp/test", O_WRONLY);
     write(fd, "Hello, World!", 13);
     close(fd);
\end{verbatim}
The first line opens the file for writing, yielding a descriptor \texttt{fd}.
Read and write operations take a memory buffer and a length
specifying how many bytes to transfer from/to this buffer.
In the example, the buffer contains the string ``Hello, World!'' with a length
of 13 bytes. The VFS keeps track of the current read/write offset into the file.
Access is sequential and writes at the end of the file implicitly increase its size.

The VFS state consists of a registry $\oh$ of open file handles, formalized as:
\begin{gather*}
\textbf{state var } \oh :\ \N \pto \Handle \qquad \text{where }\\
\textbf{data }\ \Handle = \mkhandle(\sino: \Ino, \spos : \N, \smode : \Mode)
\end{gather*}
We have proved the following invariant on the VFS state. For all $\fd$:
\begin{align}
\fd \in \oh \to \oh[\fd].\sino \in \files
\end{align}
The VFS client may obtain a file descriptor $\fd : \N$ by the call to
\texttt{vfs\_open(path, mode; fd, err)} and release it after use with
\texttt{vfs\_close(fd; err)}.
The $\spos$ slot of a handle can be modified by the operation
\texttt{vfs\_seek(fd; n, whence, err)}, returning the updated position
(the parameter \texttt{whence} specifies whether \texttt{n} is to be understood
relative to $\spos$ or absolute).

The signatures of read and write are \texttt{vfs\_read(fd; buf, len, err)} and
\texttt{vfs\_write(fd, buf; len, err)}, where buffers \texttt{buf} are arrays of
bytes ($\textbf{type } \Buffer = \Array\langle\Byte\rangle$)

\subsection{Read \& Write}
\label{sec:rw:io}

Reading and writing maps a linear buffer onto the file's array of pages.
There is a number of special cases that must be considered.

\begin{figure}
    \parbox{0.5\textwidth}{
        \centering
        \includegraphics[height=2cm]{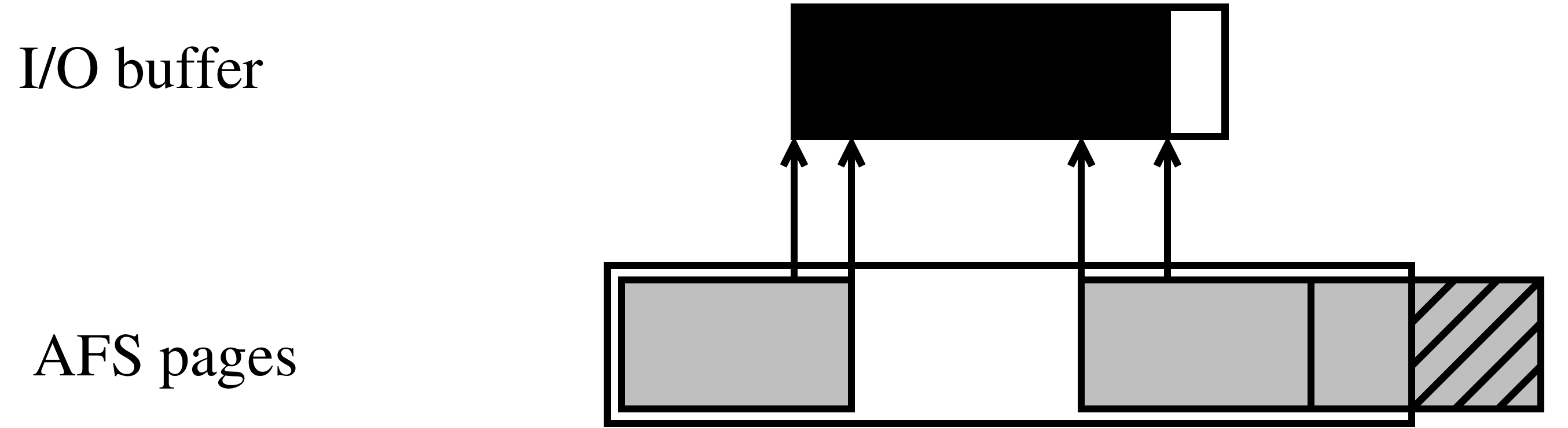}
        \caption{Read with unallocated pages}
        \label{fig:read}
    }
    \parbox{0.5\textwidth}{
        \centering
        \includegraphics[height=2cm]{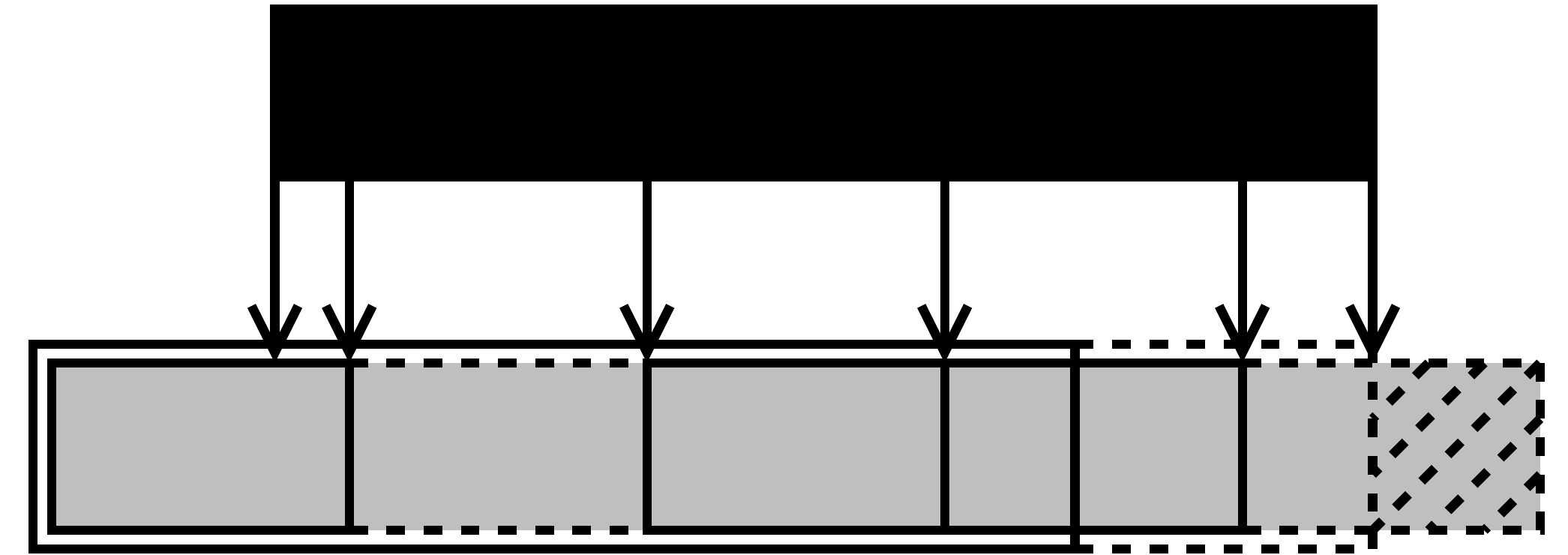}
        \caption{Write that extends the file size}
        \label{fig:write}
    }
\end{figure}

An example read operation is visualized in \Fig{read}.
At the bottom, the file's pages are denoted by gray boxes with an outer frame
indicating the file size. The hatched part of the last page does not contribute
to the file's content and must always contain zeros. The white space among the
pages denotes an unallocated page, which implicitly represents a range of all
zeros in the file.

The rectangle in the middle denotes the destination buffer (parameter
\texttt{buf}). The black part corresponds to the range that should be read
(parameter \texttt{len}). The buffer may be larger than \texttt{len} (white part).
The read operation loads the affected pages sequentially and copies the required
parts (graphically delimited by arrows) into the buffer.

Listing~\ref{fig:read-block} shows the core helper procedure that is called in a
loop until done or an error occurs. Parameters \texttt{start} and \texttt{end}
define the range to read, as absolute positions into the file in bytes.
Of this range, \texttt{total} bytes have been processed so far
(note that \texttt{done}, \texttt{buf} and \texttt{total} are passed by reference).

The procedure computes the current page and offset into that page and considers
three upper bounds for the length of the range to copy in this iteration:
the maximum number of bytes to transfer (\texttt{len}),
the end of the current page, and the end of the file.
Note that for a top-level read operation, each page is loaded at most once.
Nonexistent pages are handled in AFS:
\begin{lstlisting}[style=kiv,language=kiv,mathescape=true]
afs_readpage(ino, pageno; page, err)
  let pages = files[ino].pages in
    page := if (pageno _in pages) then pages[pageno] else emptypage
    err  := ESUCCESS
\end{lstlisting}

Writing is done similarly, an example is shown in \Fig{write}, assuming the
operation is executed with the same file as in \Fig{read}.
A write operation may extend the file at the end. In this case, the dotted lines
indicate the newly allocated parts of the file: the missing second page is
written, as well as an additional page at the end.
The fourth page is overwritten and becomes part of the file entirely.
The write operation relies on a helper routine similar to
\texttt{vfs\_read\_block}, an excerpt is shown in \Fig{write-block}.
The affected page is loaded, modified and then written back.%
\footnote{The \texttt{afs\_readpage} could be optimized away if \texttt{n = PAGE\_SIZE}}
Writes that extend the file additionally call \texttt{afs\_truncate} with the
new size.

A write that completely falls beyond file size leads to a gap in the file
that must contain zeros. That no zeros have to be written in between is
guaranteed by the invariants \Eqn{afs:pages} and \Eqn{afs:zero}.

A further detail is that a read/write may be successful, even if \emph{less} than
\texttt{len} bytes were transferred, i.e., some intermediate page access failed,
for example if the storage medium becomes full.

\begin{figure}[t]
    \centering
\begin{lstlisting}[style=kiv,language=kiv,mathescape=true]
vfs_read_block(start, end, inode; done, buf, total, err)
  if   _not done && start + total <= inode.size
     && total <= # buf && inode.ino _in files
     && err = ESUCCESS then
  let pageno = (start + total) / PAGE_SIZE        ;; integer division
      offset = (start + total) % PAGE_SIZE        ;; and     modulo
      page   = ?
  in ;; bytes to read in this iteration
     let n = min(end        - (start + total)     ;; read size    boundary
                 PAGE_SIZE  -  offset             ;; current page boundary
                 inode.size - (start + total)) in ;; inode size   boundary
     if n != 0 then
       afs_readpage(inode.ino, pageno, dirs, files; page, err)
       if err = ESUCCESS then
         buf   := copy(page, offset, buf, total, n)
         total := total + n
     else done := true
\end{lstlisting}
    \caption{ASM rule to read a partial page}
    \label{fig:read-block}
\begin{lstlisting}[style=kiv,language=kiv,mathescape=true]
     if n != 0 then
       afs_readpage(inode.ino, pageno, dirs, files; page, err)
       if err = ESUCCESS then
         page := copy(buf, total, page, offset, n)
         afs_writepage(inode.ino, pageno, page, dirs; files, err)
\end{lstlisting}
    \caption{ASM rule to write a partial page (excerpt)}
    \label{fig:write-block}
\end{figure}

\subsection{Deletion}
\label{sec:evict}

The POSIX standard allows file handles to point to files that are not part of
the directory tree any more. This situation occurs, when the last link to
an open file is removed. The manual of \texttt{close(3)} specifies:

\textsl{``If the link count of the file is 0, when all file descriptors
associated with the file are closed, the space occupied by the file shall be
freed and the file shall no longer be accessible.''}

Some applications actively exploit this feature to create hidden temporary files
(MySQL caches, Apache SSL mutex).
Replacement of files during a system update is another use case:
existing files, in particular application binaries, are unlinked before the new
version is written.
The command \texttt{lsof +L1} can be used to detect applications that still refer to a
deleted file.

Since the AFS layer does not know about open files (it can not access $\oh$),
VFS needs to signal explicitly when files become obsolete,
which is done by a call to the operation \texttt{afs\_evict}.
The helper routine \texttt{vfs\_putinode} is called after a file handle has been
closed and after a link to a file has been removed.
\begin{lstlisting}[style=kiv,language=kiv,mathescape=true,multicols=2]
vfs_putinode(ino; err)
  if _not is-open(ino, oh)
  then afs_evict(ino; err)

afs_evict(ino; err)
  if links(ino, dirs) = []
  then files := files -- ino
  err := ESUCCESS
\end{lstlisting}
where \quad
$ \isopen(\ino, \oh) \
    :\leftrightarrow \
        \exists\ \fd \in \oh.\ \ \oh[\fd].\sino = \ino$

As previously indicated, \texttt{afs\_evict} must always succeed.
The reason is that the state has already been modified and thus the
corresponding VFS call may not fail any more (see \Sec{vfs:operations}).
The requirement not to fail is reasonable, since \texttt{afs\_evict} will be
implemented as in-memory operation.%
\footnote{Technically, in UBIFS, the modification has already been recorded in
the on-flash journal and \texttt{evict} only removes the file and its content
from the RAM index.}.
The Linux VFS has a similar requirement (\texttt{void} return type of \texttt{evict}).

\section{Related Work}
\label{sec:related}

File system correctness has been an active research topic for some time.
An early model of the POSIX standard written in Z by Morgan and Sufrin is \cite{MorganSufrin87}.
Several mechanized models have been developed related to this work
\cite{ButterfieldWoodcock07,Hesselink-Lali-FACS12,ButlerAbrial08filesystem,DamchoomButler09write}.
These approaches typically remain on a very high abstraction level,
make strong simplifications, e.g., leave out hard-links or treat file content atomically.
To our knowledge, the separation of common functionality (VFS) versus file
system specific parts (AFS) has not been addressed previously.

The work of Kang and Jackson \cite{kang-jackson-flash-alloy08} is closest to our
work with respect to read and write -- it provides the same interface
(buffer, offset, length). However, their model only deals with file content but
not with directory trees or file handles. They check correctness with respect
to an abstract specification for small bounded models.
Kang and Jackson address further issues as well (fault tolerance and physical
disk layout), that can be modularly realized as a separate layer in our
refinement chain.
In comparison, their read and write algorithm is less practical than ours,
because it relies on an explicit representation of a list of blocks that needs
to be modified during an operation.

Arkoudas et al. \cite{ArkoudasKuncak-ICFEM04} address reading and writing of
files in isolation (without file handles).
Their model of file content is similar to ours (i.e., non-atomic pages).
They prove correctness of read and write with respect to an abstract POSIX-style
specification.
However, their file system interface allows only to access \emph{single bytes} at
a time, which is a considerable simplification.

Damchoom et al. \cite{ButlerAbrial08filesystem} start
with a graph-based specification of file system operations, with an interface
similar to our AFS layer. Their model differs from the VFS data model by using
parent pointers to encode the graph.
In \cite{DamchoomButler09write} Damchoom et al. decompose the write operation
with respect to pages, however, not down to bytes. They use shadow copies of
whole files to achieve abstract fault tolerance, which is not realistic.

Ferreira et al. \cite{Ferreira08} et. al. provide a POSIX-like specification 
at the level of paths.

\section{Discussion and Conclusion}
\label{sec:conclusion}

We have presented a formal model of a Virtual Filesystem Switch and an abstract
specification of its internal interface, inspired by the implementation of the
POSIX file system interface in Linux.

Our model has two notable simplifications that are visible in the external interface:
Symbolic links are not supported since we feel that they are a secondary concern.
In POSIX, \texttt{readdir} is specified only by a C library interface
based on further data structures that out of scope of this model.
We therefore chose a simple implementation of \texttt{readdir}.
In contrast to the corresponding Linux system call, it is not based on directory
handles and returns all entries at once.

The effort to develop these models was dominated by two factors:
On one hand, we spent much time studying the Linux source code,
figuring out the interplay between VFS and file system implementations and
details of the semantics of operations (e.g., evict).
On the other hand, we had to ensure that the simulation proofs involving POSIX,
the existing UBIFS model, and further refinements will work out
(e.g., nondeterministic errors).
The development of a POSIX model has been overlapped with this work in order
to clarify the requirements to the VFS.

We estimate that the effort related to VFS and AFS was about four months,
of which the technical aspects
-- writing ASM rules, specifying and proving invariants and properties --
took roughly one month.

The modularization into VFS and AFS allows us to focus on the FFS internals in
our future work.

Three important orthogonal aspects remain for future work:

A major decision was that the VFS layer does not store or cache any inode,
dentry or page objects internally. We expect that caching can be introduced by a
refinement of the VFS model without the need to change AFS.

Concurrency is an essential part of the Linux VFS. It leads to locking and
synchronization and introduces additional complexity. 
Work in this direction could benefit from Galloway et al. \cite{Galloway-VFSmodel09},
where Linux VFS code is abstracted to a SPIN model to check correct usage of
locks and reference counters.

Fault tolerance against power loss is of great interest and we are currently
proving that the model can deal with crashes anytime during the run of an
operation, using the temporal program logic of KIV \cite{ITL-TIME-2011}.

\bibliographystyle{eptcs}
\bibliography{references}

\end{document}